# Crystalline water intercalation into the Kitaev honeycomb cobaltate $Na_2Co_2TeO_6$

Masaaki Ito[1], Yuya Haraguchi[1,*], Teruki Motohashi[2], Miwa Saito[2], Satoshi Ogawa[2], Takashi Ikuta[3], and Hiroko Aruga Katori[1]

[1]*Department of Applied Physics and Chemical Engineering, Tokyo University of Agriculture and Technology, Koganei, Tokyo 184-8588, Japan*
[2]*Department of Applied Chemistry, Faculty of Chemistry and Biochemistry, Kanagawa University, Yokohama 221-8686, Japan*
[3]*Department of Biomedical Engineering, Tokyo University of Agriculture and Technology, Koganei, Tokyo 184-8588, Japan*
*Correspondence author: <u>chiyuya3@go.tuat.ac.jp</u>

We herein report the successful intercalation of water molecules into the layered honeycomb lattice of $Na_2Co_2TeO_6$, a Kitaev-candidate compound, to obtain the hydrated phase $Na_2Co_2TeO_6 \cdot yH_2O$ ($y \approx 2.4$). Fourier transform infrared spectroscopy, thermogravimetric analysis, differential scanning calorimetry, and Rietveld refinements indicate that crystalline water resides between the cobalt-based honeycomb layers. This insertion of neutral molecules significantly alters the crystal structure, increasing the interlayer spacing and modifying the local bonding environment. Magnetization measurements reveal an antiferromagnetic transition at $T_N \approx 17.2$ K, accompanied by a discernible weak ferromagnetic component. The application of moderate magnetic fields induces a spin-flop reorientation at $\mu_0 H \approx 5.7$ T. The $\lambda$-type anomaly and long-range order persist up to 9 T, showing the reconfiguration of the ground state as opposed to its suppression. Heat-capacity analysis reveals the full $2R\ln 2$ magnetic entropy expected for two $J_{eff} = 1/2$ moments per formula unit, confirming the pseudospin description. These findings demonstrate that water intercalation is a robust strategy for tuning the magnetic properties of honeycomb lattice materials. Overall, this study highlights neutral-molecule insertion as a promising route toward the discovery and engineering of quantum magnets based on layered transition metal oxides.

## I. INTRODUCTION

The discovery of unconventional charge-neutral quasiparticle excitations in quantum many-body systems has recently emerged as a central issue in condensed matter physics [1–3]. Quantum spin liquids (QSLs) are of particular interest because they exhibit no long-range magnetic order, even at temperatures near absolute zero, although they retain liquid-like spin fluctuations [4–7]. In most spin systems, long-range magnetic order emerges in the ground state. However, under sufficiently strong frustration, the magnetic order is suppressed and the spin degrees of freedom can become "fractionalized," behaving as independent quasiparticles. Although numerous theoretical approaches have been proposed to elucidate and realize QSL states, the Kitaev model has garnered particular attention because it hosts a QSL ground state and can be exactly solved [8, 9].

The Kitaev model is characterized as a spin-1/2 system with a honeycomb lattice, in which bond-dependent Kitaev interactions represent a paradigmatic example of a QSL [10]. More recently, attention has extended beyond low-spin $d^5$ systems to include high-spin $d^7$ candidates. In such systems, the interplay between $t_{2g}$ and $e_g$ orbitals—and their relationship to lattice distortions—gives rise to alternative avenues for Kitaev physics [11, 12]. However, despite intense theoretical and experimental efforts, most candidate compounds still exhibit magnetic ordering at finite temperatures, instead of forming QSLs [13–45]. The following two main factors appear to limit the formation of Kitaev QSLs: (1) competition among several exchange interactions—including isotropic Heisenberg and off-diagonal anisotropic terms—which are comparable in magnitude to Kitaev interactions, and (2) local structural distortions and orbital hybridization modifications, which reduce the effective spin–orbit-driven Kitaev term [46].

These challenges are particularly severe in high-spin $d^7$ systems. In these materials, the superexchange pathways involve both $t_{2g}$ and $e_g$ orbitals, resulting in complex exchange mechanisms. Furthermore, local distortions such as trigonal distortions often affect the interorbital interactions, making it difficult for the Kitaev interactions to dominate. Consequently, local structural distortions must be effectively and selectively controlled to realize Kitaev physics. Conventional routes, such as chemical substitution and high-pressure synthesis, have been explored extensively; however, they have not produced a QSL phase.

A promising solution, occasionally referred to as "supra ceramics," involves intercalating molecular units into an inorganic crystal framework to modify the local structure [47]. Water ($H_2O$) molecules are especially attractive for this purpose because they are electrically neutral, thereby minimizing changes in the valence state, and they offer rich local degrees of freedom through hydrogen bonding. These features enable selective local distortions that are difficult to achieve through traditional chemical or physical pressure.

$Na_2Co_2TeO_6$ is a layered honeycomb Mott insulator in which $Co^{2+}$ ($3d^7$) ions form an edge-sharing octahedral network that orders in a zigzag-type antiferromagnetic structure below $T_N \approx 27$ K [23]. Magnetic susceptibility measurements yield an effective moment of $\mu_{eff} \approx 5.64\ \mu_B$ per Co, which is consistent with high-spin $S = 3/2$ ions that map onto $J_{eff} = 1/2$ pseudospins under trigonal compression and spin–orbit coupling [23]. First-principles calculations show that trigonal distortions split the Co $t_{2g}$ manifold and, together with strong spin–orbit coupling, give rise to an effective $J_{eff} = 1/2$ ground state with significant XY-type anisotropy [16, 19, 48]. Such a pseudospin basis is necessary for Kitaev-type bond-dependent exchange; however, in $Na_2Co_2TeO_6$ this Kitaev term competes with sizable isotropic Heisenberg ($J_1$, $J_3$) and off-diagonal $\Gamma$ terms of comparable magnitude [20]. Because the pristine electronic structure features this delicate balance of competing exchanges and orbital anisotropies, further local distortions—such as those induced by water intercalation—can tip the system closer to or farther from the Kitaev QSL regime [11−13].

On the basis of these insights into the pristine electronic and magnetic properties of $Na_2Co_2TeO_6$, we investigated how controlled $H_2O$ intercalation perturbs this balance and modifies the structural framework and resulting magnetic order. Our initial findings showed that this material spontaneously incorporates water under ambient conditions, leading to partial expansion of the interlayer spacing. However, the hydrated phase fraction produced under these conditions was minimal, which hindered systematic structural investigation. Therefore, we developed a method for targeted synthesis of the hydrated phase, $Na_2Co_2TeO_6 \cdot yH_2O$, which involved stirring the parent compound in warm water under carefully controlled conditions. Hence, we produced a sufficient quantity for detailed measurements. We used X-ray diffraction, thermogravimetric analysis, and magnetic susceptibility studies to investigate how $H_2O$ intercalation distorts the $CoO_6$ octahedra and the effects on the Kitaev and other spin-exchange interactions. We assessed these changes based on the variations in the antiferromagnetic transition temperature and Weiss temperature.

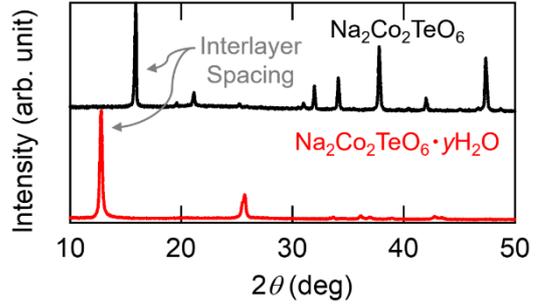

**Figure 1.** XRD patterns of $Na_2Co_2TeO_6$ before (top, black) and after (bottom, red) immersion in water at 60 °C for 7 d under stirring. After immersion, the peaks corresponding to the interlayer spacing shifted toward lower angles, indicating an increase in the interlayer distance.

## II. EXPERIMENTAL METHODS

The precursor $Na_2Co_2TeO_6$ was prepared using a conventional solid-state reaction method [14]. Specifically, $Na_2CO_3$ (FUJIFILM Wako Pure Chemical Corporation, 99.8%), CoO (Kojundo Chemical Lab. Co., Ltd., 99.7%), and $TeO_2$ (Kojundo Chemical Lab. Co., Ltd., 99.9%) were weighed out with a molar ratio of 1.05:2:1 and then heated in air at 900 °C for 96 h. The resulting polycrystalline $Na_2Co_2TeO_6$ powder was sealed in a Teflon-lined stainless-steel autoclave with distilled water and heated at 60 °C for 7 d under stirring to achieve hydration.

Powder X-ray diffraction (XRD) measurements before hydration, after hydration, and after dehydration at 200 or 300 °C for 12 h were performed using Cu-K$\alpha$ radiation. Rietveld refinements and maximum entropy method (MEM) analyses were conducted using the Z-RIETVELD software package [49].

Fourier-transform infrared (FT-IR) spectroscopy was used to verify the intercalation of $H_2O$ and to distinguish between intact water molecules and hydroxyl groups. The FT-IR measurements were conducted using an FT/IR-4700 (JASCO) with the single attenuated total reflection method (JASCO, ATR PRO 650G). Additional spectra were collected from $CoSO_4 \cdot 7H_2O$ and $Co(OH)_2$ as references. Thermogravimetric analysis (TGA) under a nitrogen atmosphere coupled with quadrupole mass spectrometry (QMS) was used to detect and identify the gaseous products that evolved upon heating. The ion signals were monitored at their corresponding mass-to-charge ratios $m/z$. The TGA data were acquired using a Thermo Plus EVO2 TG-DTA8122 (Rigaku) and the QMS measurements were conducted using a Transpector CPM (Inficon).

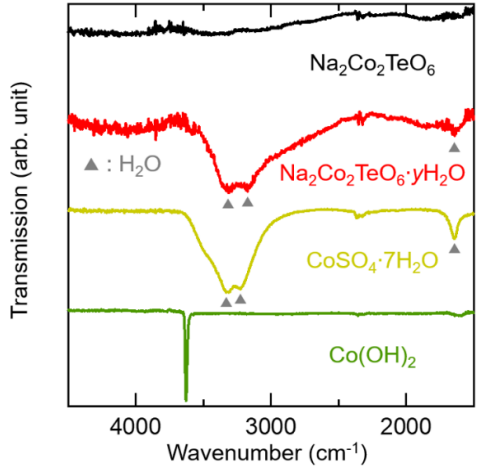

**Figure 2.** FT-IR spectra of $Na_2Co_2TeO_6$ before (black) and after (red) immersion in water at 60 °C for 7 d under stirring. The spectra of $CoSO_4 \cdot 7H_2O$ (yellow) and $Co(OH)_2$ (green) are also shown for comparison. The upward triangle markers indicate absorption bands originating from $H_2O$.

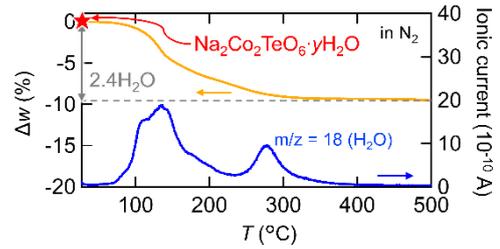

**Figure 3.** TGA and QMS results for $Na_2Co_2TeO_6 \cdot yH_2O$ under an $N_2$ atmosphere. The orange curve indicates the mass change during the heating process, while the blue curve shows the ion current for $m/z = 18$, which was attributed to the desorption of $H_2O$.

The magnetic properties were measured using a magnetic property measurement system (MPMS; Quantum Design) at the Institute for Solid State Physics (ISSP), the University of Tokyo. The temperature-dependent magnetic susceptibilities were measured under magnetic fields between 10 mT and 7 T, and isothermal magnetization curves were recorded at temperatures between 2 and 30 K. The temperature dependence of the heat capacity was measured using the conventional relaxation method in a physical property measurement system (PPMS; Quantum Design) at ISSP, the University of Tokyo.

## III. RESULTS

### A. Crystal structure

Figure 1 shows the XRD patterns for $Na_2Co_2TeO_6$ before and after immersion in water at 60 °C for 7 d under stirring. After treatment, the 00$l$ Bragg peaks shifted toward lower angles, indicating an increase in the interlayer spacing from ~5.61 to ~6.95 Å. This shift is consistent with related findings in other honeycomb-layered hydrates such as $Na_{0.85}Co_2SbO_6 \cdot 1.7H_2O$ [15] and $NaNi_2BiO_{6-\delta} \cdot 1.7H_2O$ [50]. Similar trends have also been observed in triangular-layered hydrates such as $Na_xCoO_2 \cdot yH_2O$ [51,52], $Na_{0.3}NiO_2 \cdot 0.7H_2O$ [53], $Na_{0.22}RuO_2 \cdot 0.45H_2O$ [54], and $Na_{0.3}RhO_2 \cdot 0.6H_2O$ [55], as well as in various hydrated alkaline-intercalated transition metal dichalcogenides [56–65]. These parallels suggest a common structural response to hydration across different layered materials. In these compounds, water intercalation increases the interlayer gap by approximately 1.4 Å, which closely matches the expansion observed for $Na_2Co_2TeO_6$. Therefore, these results support the conclusion that water molecules were inserted between the layers to form $Na_2Co_2TeO_6 \cdot yH_2O$.

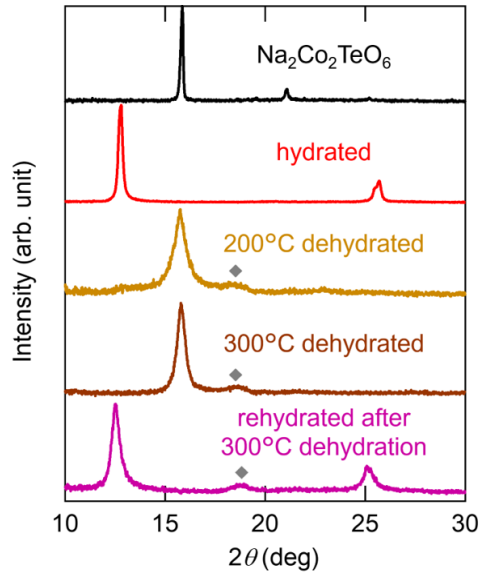

**Figure 4.** XRD patterns before hydration ($Na_2Co_2TeO_6$, black), after hydration ($Na_2Co_2TeO_6 \cdot yH_2O$, red), after dehydration at 200 °C (ochre), after dehydration at 300 °C (brown), and after rehydration of the sample dehydrated at 300 °C (purple). The diamond markers highlight peaks attributed to impurities, which likely formed during dehydration.

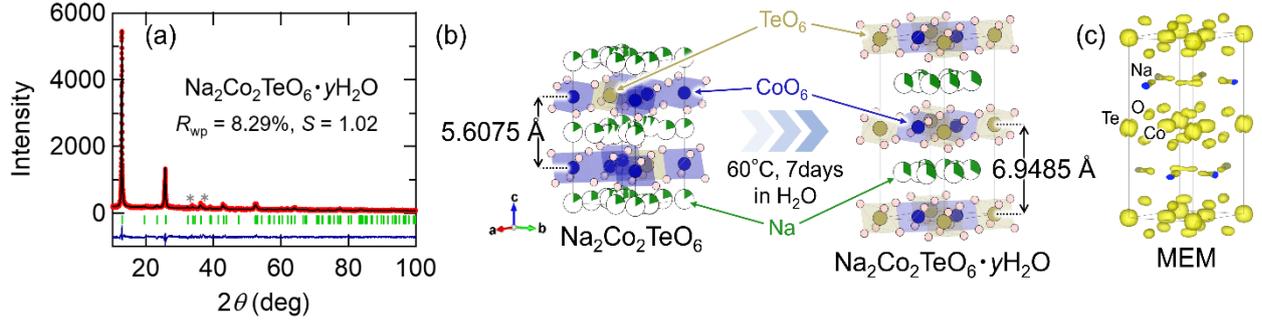

**Figure 5.** (a) Rietveld refinement of $Na_2Co_2TeO_6 \cdot yH_2O$. The observed intensities, calculated intensities, and their difference are shown in red, black, and blue. Green vertical bars indicate the positions of the Bragg reflections. Asterisks mark the impurity peaks. (b) Schematic showing the structural changes in $Na_2Co_2TeO_6$ before and after hydration. The crystal structures were visualized using the VESTA program [69]. (c) Three-dimensional electron density distribution of $Na_2Co_2TeO_6 \cdot yH_2O$, calculated via MEM.

The presence of intercalated water was confirmed by recording FT-IR spectra for pristine and hot-water-treated $Na_2Co_2TeO_6$ (Figure 2). The treated sample exhibited distinct absorption peaks at approximately 3400 and 1600 cm$^{-1}$, corresponding to the stretching and bending modes of $H_2O$, respectively [66]. These features also appeared in the reference compound $CoSO_4 \cdot 7H_2O$, reinforcing the conclusion that $H_2O$ was incorporated into the crystal structure. No discernible bands attributable to OH groups were observed, indicating that the $H_2O$ molecules remained intact rather than being converted into OH bonds. This was further corroborated by comparison with $Co(OH)_2$, which showed a strong peak at 3600 cm$^{-1}$ arising from the OH groups [67]. Hence, in the hot-water-treated $Na_2Co_2TeO_6$, water was intercalated as molecular $H_2O$, which expanded the interlayer and modified the magnetic interlayer coupling.

Figure 3 shows the TGA/QMS analysis of $Na_2Co_2TeO_6 \cdot yH_2O$. Stepwise weight loss was accompanied by a detectable ion current at m/z = 18, which corresponds to the release of $H_2O$ up to 500 °C. From the total weight loss, the water content before heating was estimated to be $y = 2.4$. The two distinct peaks in the QMS data indicate the presence of two inequivalent water sites, consistent with the structural model described below.

Figure 4 shows the changes in the XRD profile of $Na_2Co_2TeO_6 \cdot yH_2O$ after heat treatment at 200 and 300 °C, which indicates that the interlayer spacing was approximately the same as that of the pristine phase.

**Table 1.** Crystallographic parameters of $Na_2Co_2TeO_6 \cdot yH_2O$ determined from powder XRD data. The structure was refined in the space group $P6_3/mcm$, with lattice constants $a = 5.3160(12)$ Å and $c = 13.897(3)$ Å. $B$ denotes the atomic displacement parameter. WO denotes the oxygen in the $H_2O$ molecule.

| Atom | Site | Occupancy | x | y | z | B (Å) |
|---|---|---|---|---|---|---|
| Co | 4d | 1 | 2/3 | 1/3 | 0 | 0.92294(15) |
| Te | 2b | 1 | 0 | 0 | 0 | 0.078459(13) |
| Na1 | 4c | 0.01(2) | 2/3 | 1/3 | 1/4 | 1.3212(2) |
| Na2 | 12j | 0.33(4) | 0.588(2) | -0.173(5) | 1/4 | 1.3212(2) |
| O | 12j | 1 | 0.690(3) | 0.690(3) | 0.57260(15) | 0.48566(8) |
| WO1 | 2a | 0.32(4) | 0 | 0 | 1/4 | 1.3212(2) |
| WO2 | 12j | 0.34(4) | 0.588(2) | -0.173(5) | 1/4 | 1.3212(2) |

This demonstrates the reversibility of the hydration reaction. This process reflects the release of crystalline water, in line with the TGA/QMS results. However, the interlayer spacings at 200 and 300 °C (~5.62 and ~5.60 Å, respectively) were slightly offset from the original value (~5.61 Å). Moreover, impurities appeared to form during dehydration and persisted after rehydration, which yielded a hydrated phase with an interlayer spacing of 7.089 Å. Comparing the magnetic susceptibilities of the dehydrated and prehydrated samples (see Supplementary Information [68]) indicated that the hydration–dehydration process was not fully reversible. Partial structural collapse and/or Na removal may have accompanied these cycles, affecting the purity and structural integrity of the sample. Despite these complications, the overall observations confirm the formation of $Na_2Co_2TeO_6 \cdot yH_2O$ after immersing $Na_2Co_2TeO_6$ in hot water under stirring.

Figure 5(a) compares the fitted and observed diffraction patterns from the Rietveld refinement of $Na_2Co_2TeO_6 \cdot yH_2O$. The major Bragg reflections were indexed to the $P6_3/mcm$ space group, analogous to that of $K_2Co_2TeO_6$ [35], rather than the $P6_322$ symmetry of pristine $Na_2Co_2TeO_6$ [14]. A structural model was generated by substituting Na for K in the $K_2Co_2TeO_6$ framework and introducing oxygen atoms corresponding to water molecules within the Na layer. Following initial refinement, MEM was used to construct an electron density map and refine the positions and site occupancies of Na and water. Iterative feedback between the Rietveld refinement and MEM yielded the final structural model shown in Figure 5(b), which had reliability factors $R_{wp}$ = 8.29% and $S$ = 1.02. The refined structural parameters are summarized in Table 1.

Two distinct crystallographic sites for water, WO1 and WO2, were identified in the final model, corroborating the two $H_2O$ release processes observed using QMS. Both water sites were located in the same plane as the partially occupied Na sites. The WO2 site was shared with Na2 (each having an occupancy of approximately 0.34 and 0.33, respectively), indicating a high degree of disorder similar to that found in pristine $Na_2Co_2TeO_6$ [14]. The interlayer distance increased substantially from 5.6075 Å in the anhydrate phase to 6.9485 Å after hydration, which enhanced the two-dimensional character of the magnetic layers.

A final round of MEM calculations using the converged structural parameters provided the electron density distribution depicted in Figure 5(c). The spatial extent and intensity of the electron density were consistent with the refined structure, confirming the reliability of the model. Bond valence sum [70] analysis showed that the Co ions had an oxidation state of +2.043, which is consistent with a formal $Co^{2+}$ state. Collectively, these findings confirm that water intercalation in $Na_2Co_2TeO_6$ increased the interlayer gap and potentially affected the magnetic interactions between the layers.

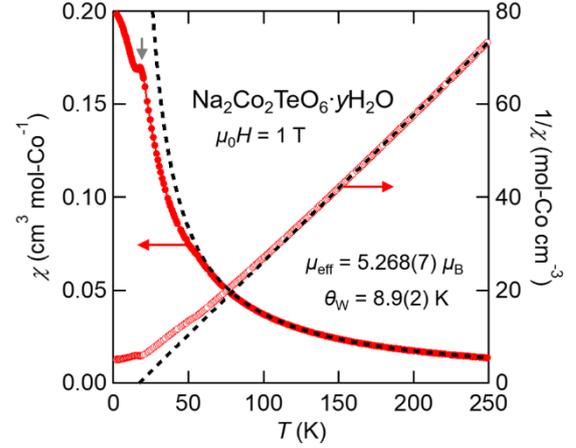

**Figure 6**. Temperature dependence of the magnetic susceptibility $\chi$ and its inverse $1/\chi$ for a polycrystalline sample of $Na_2Co_2TeO_6 \cdot yH_2O$ under an applied magnetic field of $\mu_0 H$ = 1 T. The black dashed line indicates the Curie–Weiss fit. The magnetic anomaly observed at approximately 16 K is indicated by the vertical arrow.

### B. Magnetic properties

Figure 6 shows the temperature dependence of the magnetic susceptibility $\chi$ and its inverse $1/\chi$ for a polycrystalline sample of $Na_2Co_2TeO_6 \cdot yH_2O$ under an applied field of $\mu_0 H$ = 1 T. A Curie–Weiss fit for $1/\chi$ between 150 and 250 K yielded an effective magnetic moment of $\mu_{eff}$ = 5.268(7) $\mu_B$ and a positive Weiss temperature $\theta_w$ = 8.9(2) K. This $\mu_{eff}$ exceeded the spin-only contribution expected for $Co^{2+}$ (3.87 $\mu_B$), which suggests that unquenched orbital angular momentum contributed significantly to the magnetic moment. Large effective moments exceeding the spin-only value have been observed in other $Co^{2+}$-based magnets, including the honeycomb-lattice compounds $Na_2Co_2TeO_6$ [16], $Na_3Co_2SbO_6$ [17], and $Li_3Co_2SbO_6$ [18], as well as in $Na_2BaCo(PO_4)_2$ [71]. Two distinct anomalies emerged in $\chi$ and $1/\chi$. The first anomaly at approximately 16 K was consistent with antiferromagnetic ordering, whereas the second anomaly at approximately 48 K was attributed to trace amounts of adsorbed oxygen.

The behavior at approximately 16 K was clarified by measuring the temperature dependence of $\chi$ under

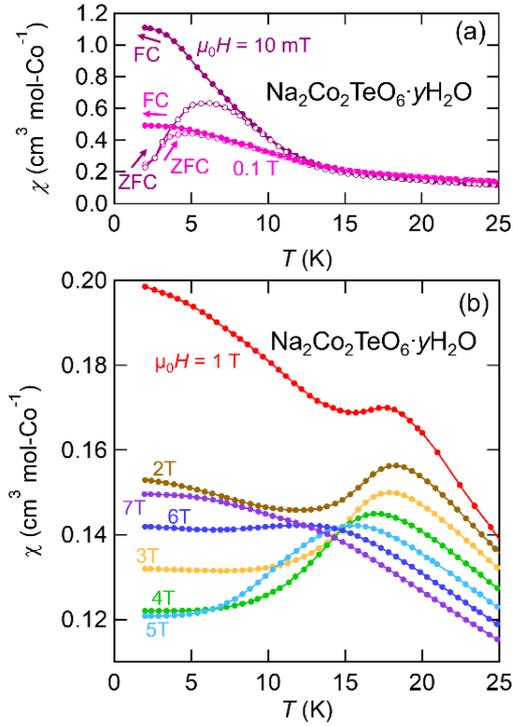

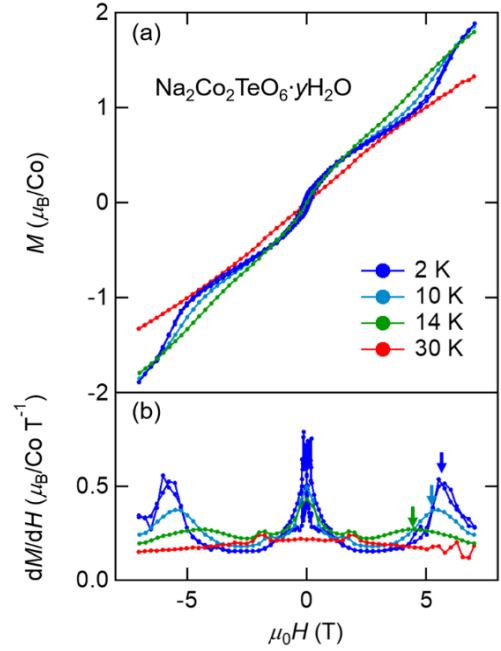

**Figure 7.** (a) Temperature dependence of the magnetic susceptibility $\chi$ of $Na_2Co_2TeO_6 \cdot yH_2O$ under applied magnetic fields of $\mu_0H = 10$ mT and 0.1 T, measured in the FC (filled circles) and ZFC (open circles) modes. The onset of the ferromagnetic-like increase in $\chi$ is marked by a downward triangle. (b) Temperature dependence $\chi$ under various magnetic fields measured in the FC mode.

applied fields between 10 mT and 7 T, as shown in Figures 7(a)–7(c). At 10 mT and 0.1 T, the zero-field-cooled (ZFC) and field-cooled (FC) susceptibility curves diverged. An additional feature, indicated by the downward triangles in Figure 7(a), shows that $\chi$ was enhanced, which persisted up to fields above 1 T (Figure 7(b)). In the $\chi$ measurements, the antiferromagnetic-like behavior became more pronounced as the applied field increased to approximately 3 T. However, this feature broadened abruptly at approximately 6 T and appeared to vanish at approximately 7 T. We attributed this evolution to a field-induced spin-flop transition that generated a spontaneous canted moment, thereby masking the usual downturn of the antiferromagnetic susceptibility rather than eliminating the ordered state itself, which was corroborated by the specific-heat results presented below. Therefore, the anomaly at approximately 17 K marks the onset of antiferromagnetic order, whereas the high-field evolution of $\chi$ reflects the emergence of a spin-flop

**Figure 8.** (a) Isothermal magnetization curve for $Na_2Co_2TeO_6 \cdot yH_2O$ measured at various temperatures. (b) The $dM/dH$ data are presented as a function of the magnetic field, and the observed peak is indicated by a vertical arrow.

phase carrying a weak-ferromagnetic component, not the suppression of antiferromagnetism.

The presence of a weak ferromagnetic moment is likely due to the Dzyaloshinskii–Moriya (DM) interaction [72,73]. In a perfect honeycomb lattice with only the nearest-neighbor bonds, inversion symmetry at the bonds prevents the formation of a finite DM vector. However, second- or third-neighbor interactions can break the local inversion symmetry and allow the formation of a DM component, which may stabilize a noncollinear spin arrangement. Weak ferromagnetism has been reported in a variety of honeycomb-lattice magnets, including Co-based systems such as $Na_2Co_2TeO_6$ [23] and $Li_3Co_2SbO_6$ [18] and other $3d^3$ [74], $5d^5$ [75,76], and $4f^1$ [77–79] honeycomb magnets. Moreover, theoretical research on $Na_2Co_2TeO_6$ [24] has shown that a triple-Q spin order, driven by competing Heisenberg, Kitaev, and off-diagonal interactions, can produce in-plane ferrimagnetism and thus a weak ferromagnetic moment. A similar mechanism involving multiple competing interactions may underlie the weak ferromagnetism in $Na_2Co_2TeO_6 \cdot yH_2O$.

Figures 8(a) and 8(b) show the isothermal magnetization curves and their derivatives at selected temperatures. A small hysteresis loop was observed at approximately 0 T, which corroborates the existence of a weak ferromagnetic component. The remanent magnetization was approximately 0.07 $\mu_B$, corresponding to approximately 4% of the expected 2 $\mu_B$ saturation magnetization for a $J_{eff} = 1/2$ state with $g \approx 4$ [80,81]. This fraction is considerably larger than the weak ferromagnetic moments previously reported for $Na_2Co_2TeO_6$ [25] and $Li_3Co_2SbO_6$ [18]. Assuming the entire sample adopts a canted antiferromagnetic order, the spin canting angle will be approximately 2°.

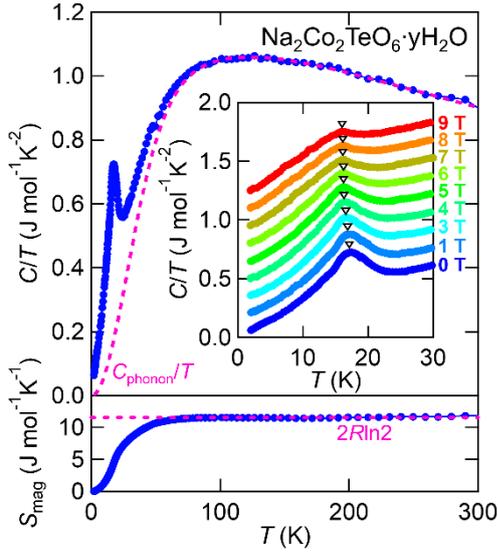

**Figure 9.** (Top) Temperature dependence of the heat capacity of $Na_2Co_2TeO_6 \cdot yH_2O$ plotted as $C/T$. The pink dashed curve represents the phonon contribution extracted from the Debye–Einstein fit [82] for the data above 80 K. (Inset) $C/T$ measured under fields of 0–9 T. Each trace is vertically offset by 0.15 J mol$^{-1}$ K$^{-2}$ for clarity. Triangles indicate $T_N$. (Bottom) Magnetic entropy $S_{mag}$ under zero magnetic field, obtained by integrating $C_{mag}/T$. The horizontal dashed line indicates $S_{mag} = 2R\ln2$, the full entropy expected for two $J_{eff} = 1/2$ moments per formula unit.

In addition to the small hysteresis, the magnetization curves at 2 K exhibited a jump at approximately $\mu_0H = \pm5.7$ T, as shown for both $M$ and $dM/dH$ (Figure 8(b)). Guided by the specific-heat results presented below, we did not interpret this jump as a metamagnetic, forced-ferromagnetic collapse of the antiferromagnetic order; instead, we considered it to be the critical field of a spin-flop transition. Above 5.7 T, the magnetization increased sharply and approached the fully-polarized 2 $\mu_B$ per Co$^{2+}$ $J_{eff} = 1/2$ moment with $g \approx 4$ [80,81]. However, the inevitable field-independent van Vleck contribution intrinsic to Co$^{2+}$ ions implied that this seemingly polarized moment could not be regarded as genuine saturation. The persistence of a sharp λ-type anomaly in $C/T$ up to 9 T (discussed in Section V. C) confirmed that the long-range antiferromagnetic order survived above 7 T, indicating that full saturation must occur under substantially higher fields. This is consistent with the canting of the $J_{eff} = 1/2$ moments in the spin-flop phase, rather than complete spin polarization. As the temperature increased, the transition field decreased and it was no longer observed at 30 K, which is consistent with a transition below $T_N$. These observations show that a modest external field reoriented the antiferromagnet into a spin-flop state without affecting the long-range order. For $dM/dH$ at 14 and 30 K, small anomalies were observed at approximately ±2 T; however, no corresponding magnetic anomalies were observed in the temperature dependence of the magnetic susceptibility or heat capacity data, as discussed later. These features may be attributed to the alignment of magnetic domains or minor contributions arising from impurities.

### C. Thermodynamic properties

Figure 9 shows the temperature dependence of the heat capacity divided by temperature $C/T$ for $Na_2Co_2TeO_6 \cdot yH_2O$. A sharp λ-type peak was observed at $T_N$, signaling the onset of magnetic order. The lattice contribution was determined by fitting the 80–300 K range with a composite Debye–Einstein phonon model [82]. Subtracting this baseline yielded the magnetic term $C_{mag}/T$ (see Supplementary Information [68]). Integrating $C_{mag}/T$ recovers the magnetic entropy $S_{mag}$ (Figure 9, bottom panel), which converges to $2R\ln2$. This is consistent with $J_{eff} = 1/2$ Kramers doublets and confirms the validity of the lattice subtraction.

The inset in Figure 9 shows $C/T$ under various applied magnetic fields. The λ-peak broadened and its maximum shifted from 17.2 K at 0 T to 15.9 K at 9 T, which identified the anomaly as an antiferromagnetic transition with a finite field derivative. The persistence of the peak at 9 T, despite the broadening, indicates that the ordered state was not destroyed by the field. Together with the disappearance of the low-temperature downturn in $\chi$ and the magnetization jump at approximately 5.7 T, this field evolution supports a spin-flop interpretation, rather than a metamagnetic collapse into a fully polarized state.

### IV. DISCUSSION

The magnetization and heat capacity data show that $Na_2Co_2TeO_6 \cdot yH_2O$ underwent an antiferromagnetic transition at $T_N \approx 17.2$ K under zero magnetic field, with a weak ferromagnetic moment generated by spin canting. The application of a magnetic field did not extinguish this order; instead, it triggered a spin-flop reorientation at $\mu_0 H \approx 5.7$ T. Simultaneously, $T_N$ was reduced slightly to 15.9 K at 9 T, demonstrating that the long-range order persisted throughout the observed field range. The delicate balance of interactions—and thus the magnetic ground state—was tuned by the intercalation of neutral water molecules between the layers.

We investigated the effects of water intercalation on the magnetic properties by first examining the structural changes induced by hydration. In $Na_2Co_2TeO_6 \cdot yH_2O$, the $c$-axis lattice parameter increased from 11.2149 to 13.8970 Å, corresponding to an increase in the interlayer spacing from 5.6075 to 6.9485 Å. This expansion is expected to enhance the two-dimensional magnetic behavior by reducing interlayer coupling. In addition, the $a$-axis showed a non-negligible increase from 5.2889 to 5.3160 Å.

Such lattice distortions affect the $CoO_6$ octahedra, which we quantified using the O-Co-O bond-angle variance $\sigma^2$ [83]. Hydration does not affect the bond valence sum of Co, i.e., its effective valence state; therefore, the changes in the $CoO_6$ octahedra can be attributed purely to crystal-field effects caused by the octahedral distortion. The parameter $\sigma^2$ is defined as

$$\sigma^2 = \sum_{i=1}^{m} \frac{(\varphi_i - \varphi_0)^2}{m-1}, \quad (1)$$

where $m = 12$ for an octahedron and $\varphi_0$ is the ideal octahedral angle of 90°. In the pristine compound ($\sigma_2$ = 60.4403 deg$^2$ [26]), the $CoO_6$ octahedra deviate moderately from perfect octahedral geometry. Upon hydration, $Na_2Co_2TeO_6 \cdot yH_2O$ shows a significantly larger $\sigma^2$ of 94.1944 deg$^2$, indicating stronger trigonal distortion. Such enhanced trigonal distortion typically promotes antiferromagnetic correlations among $Co^{2+}$ spins, consistent with the transition at $T_N \approx 15.4$ K.

As shown in Figure 10, $\sigma^2$ was compared to the magnetic transition temperature $T_N$ for various Co-based honeycomb compounds [18, 23, 28–39]. This showed a broad trend where $T_N$ increased as $\sigma^2$ increased. However, despite having a relatively large $\sigma^2$, $Na_2Co_2TeO_6 \cdot yH_2O$ ordered at ~15.4 K, which indicates that factors beyond octahedral distortion govern its magnetism.

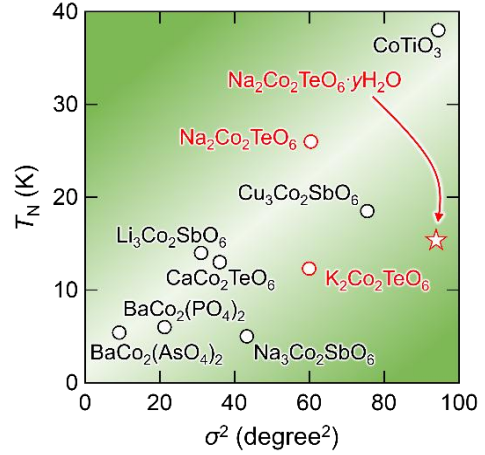

**Figure 10.** Relationship between the bond-angle variance $\sigma^2$ of the $CoO_6$ octahedra and magnetic transition temperature $T_N$ in various Co honeycomb-layered compounds. $Na_2Co_2TeO_6$, $Na_2Co_2TeO_6 \cdot yH_2O$, and the other Co honeycomb-layered materials are denoted by a red circle, a red star, and black circles, respectively. The background gradient emphasizes the observed trend that $T_N$ increases as $\sigma^2$ increases.

One key parameter is the nearest-neighbor Co–Co distance, which can change the balance between direct orbital overlap (favoring Heisenberg exchange) and more anisotropic interactions such as Kitaev or off-diagonal terms. For example, in $BaCo_2(AsO_4)_2$ and $BaCo_2(PO_4)_2$, the absence of a central cation in the honeycomb planes leads to extremely short Co–Co distances, small $\sigma^2$ values (9.1717 and 21.2611 deg$^2$, respectively), and low transition temperatures ($T_N \approx 5.4$ and 6 K, respectively) [27–31]. Neutron-scattering and first-principles calculations suggest that an XXZ model describes these materials better than a Kitaev–Heisenberg framework [32, 39, 40].

By contrast, hydration enlarges the $ab$-plane of $Na_2Co_2TeO_6$, which increases the nearest-neighbor Co–Co distance from 3.0536 to 3.0692 Å. This likely reduces direct $t_{2g}$–$t_{2g}$ overlap and weakens the conventional Heisenberg coupling, thereby increasing the importance of frustrated or anisotropic interactions. Consequently, even with pronounced trigonal distortion, the magnetic transition temperature remained relatively modest at 15.4 K. However, the ~0.5% increase in the Co–Co distance that accompanied hydration was too small to account for the ~36% drop in $T_N$. The more conspicuous structural change was the swelling of the interlayer gallery. The insertion of $H_2O$ lengthened the c-axis from ~5.61 to ~6.95 Å. As shown in Figure 10, $Na_2Co_2TeO_6 \cdot yH_2O$ and $K_2Co_2TeO_6$ [35] fell below the

$\sigma^2$–$T_N$ trend. These outliers were unified by their considerably expanded $c$ axes. This indicates that weakened interlayer exchange has an important role in depressing $T_N$. However, even this geometric decoupling cannot be the sole controlling factor, because $K_2Co_2TeO_6$—whose interlayer spacing (~6.31 Å) lies between those of anhydrous and hydrated $Na_2Co_2TeO_6$—showed the lowest ordering temperature among the samples (~4 K) despite its shorter Co–Co bond. Therefore, the comparison highlights the importance of the $c$-axis and indicates that the ordering scale is governed by the collective action of several parameters: subtle changes in the Co–Co distance, variations in the trigonal distortion of the $CoO_6$ octahedra that tune single-ion anisotropy, and the strength and coherence of interlayer coupling, which can be weakened geometrically (through layer expansion) or chemically (through alkali-ion disorder). Hydration perturbs all three parameters simultaneously, amplifying the frustration and suppressing the long-range order far more than any one adjustment in isolation. Targeted experiments that isolate individual parameters are therefore essential to determine their relative importance.

Three complementary approaches can be used to investigate the interplay between structure and magnetism in $Na_2Co_2TeO_6 \cdot yH_2O$ and the broader $A_2Co_2TeO_6$ family. First, systematic chemical or hydration tuning can be used to vary the trigonal distortion, in-plane Co–Co spacing, and interlayer separation almost independently, revealing which geometric lever most effectively stabilizes or suppresses the long-range order. Second, advanced spectroscopic techniques such as single-crystal neutron diffraction and resonant inelastic X-ray scattering can be used to track how the exchange pathways respond to determine whether anisotropic interactions or more conventional Heisenberg terms dominate in each structural regime. Third, high pressure and controlled dehydration can provide reversible control over the $ab$-plane and $c$-axis lattice parameters, enabling one-to-one correlations between the structural changes and emergent magnetic behavior.

Recent studies on thin films of the cobalt-based Kitaev candidate $Cu_3Co_2SbO_6$ have demonstrated precise control of octahedral distortions and magnetic ordering through epitaxial strain and post-growth annealing [84, 85]. However, the film geometry precludes bulk-sensitive probes such as superconducting quantum interference device (SQUID) magnetometry, heat-capacity measurements, and neutron scattering. By contrast, our water-intercalated $Na_2Co_2TeO_6 \cdot yH_2O$ method produces gram-scale, phase-pure powders whose interlayer spacing can be tuned stepwise, which are fully compatible with bulk analysis techniques. Therefore, all three experimental strategies can be applied in combination to determine quantitative links between the structural tuning and magnetic ground states. This positions hydration chemistry as a bulk analog to epitaxial strain, which promises to elucidate the delicate balance of interactions that control $T_N$ in honeycomb cobaltates, and to inform the design of two-dimensional magnets capable of hosting exotic spin-liquids or topologically non-trivial phases.

## V. SUMMARY

This study demonstrated that intercalating water into the layered Kitaev candidate $Na_2Co_2TeO_6$ yields a cobalt tellurate hydrate, $Na_2Co_2TeO_6 \cdot yH_2O$, with significant changes in the structural and magnetic properties. Hydration increases the interlayer spacing along the $c$-axis, modifies the octahedral geometry, and preserves the valence state of cobalt. The XRD, thermal analysis, and infrared spectroscopy revealed that water molecules occupy multiple interlayer sites with partial occupancies, introducing local disorder and expanding the $c$-axis (dominantly) and $a$-axis. Consequently, the antiferromagnetic transition temperature decreases from ~27 to ~17.2 K, which highlights how the delicate interplay of magnetic interactions depends sensitively on the crystal lattice. Under an applied magnetic field, the transition only shifts to 15.9 K at 9 T and a spin-flop occurs at approximately 5.7 T, which indicates that the long-range order is preserved when the spin orientation is reconfigured. Furthermore, the bond-angle variance increases significantly after water intercalation, indicating enhanced trigonal distortion of the $CoO_6$ octahedra, which typically stabilizes the antiferromagnetic order. However, the modest transition temperature suggests that additional frustration arises from the enlarged in-plane Co–Co distance and weakened direct orbital overlap. Overall, these findings indicate that water intercalation is a powerful strategy to systematically tune lattice parameters and low-temperature magnetic ground states in honeycomb cobaltates without affecting the electronic configuration of cobalt.

## ACKNOWLEDGEMENT

This work was supported by JST PRESTO "Creation of Future Materials by Expanding Materials Exploration Space" [Grant Number JPMJPR23Q8], JSPS KAKENHI Transformative Research Areas (A) "Supra-ceramics" [Grant Numbers JP23H04616, JP22H05142,

JP22H05143], JSPS KAKENHI Transformative Research Areas (A) "1000-Tesla Chemical Catastrophe" [Grant Number JP24H01613], JSPS KAKENHI Young Scientific Research [Grant Number JP22K14002], and JSPS KAKENHI Scientific Research (C) [Grant Number JP24K06953]. Part of this work was conducted as joint research at the Institute for Solid State Physics, the University of Tokyo [Project Numbers 202306-GNBXX-0128, 202311-GNBXX-0018, 202306-MCBXG-0094, and 202306-MCBXG-0070].